\ifx\undefined\documentclass
  \documentstyle[12pt,a4,axodraw]{article}
  \newcommand\mathrm[1]{{\rm #1}}
  \newcommand\mathcal[1]{{\cal #1}}
\else
  \documentclass[12pt,a4paper]{article}
  \usepackage{axodraw}
\fi

\begin{document}

\newcommand{\nl}{\nonumber \\}
\newcommand{\mc}{Monte Carlo}
\newcommand{\qmc}{Quasi-Monte Carlo}
\newcommand{\qran}{quasi-random}
\newcommand{\intl}{\int\limits}
\newcommand{\intk}{\intl_K}
\newcommand{\suml}{\sum\limits}
\newcommand{\prol}{\prod\limits}
\newcommand{\umu}{^{\mu}}
\newcommand{\order}[1]{{\mathcal O}\left(#1\right)}
\newcommand{\eqn}[1]{Eq.(\ref{#1})}
\newcommand{\intinf}{\intl_{-\infty}^{\infty}}
\newcommand{\intinfi}{\intl_{-i\infty}^{i\infty}}
\newcommand{\ddf}{{\mathcal D}f}
\newcommand{\si}{\sigma}
\newcommand{\fal}[1]{^{\underline{#1}}}
\newcommand{\bpic}{\begin{picture}}
\newcommand{\epic}{\end{picture}}
\newcommand{\Cross}[4]{\Line(#1,#2)(#3,#4)\Line(#1,#4)(#3,#2)}
\newcommand{\DL}[2]{\DashLine(#1)(#2){1.5}}
\newcommand{\DC}[2]{\DashCArc(#1)(#2,0,360){2}}
\newcommand{\BC}{\BCirc}
\newcommand{\Yv}[1]{\BCirc(#1){3}}
\newcommand{\Vix}[1]{\Vertex(#1){2}}

\pagestyle{empty}

\begin{flushright}
NIKHEF 96-02
\end{flushright}

\begin{center}
\begin{Large}
{\bf Discrepancy-based error estimates\\
     for \qmc.\\
     \vspace{\baselineskip}
     I: General formalism}\end{Large}\\
     \vspace{\baselineskip}
{\bf Jiri Hoogland${}^{1}$ and Ronald Kleiss${}^{2}$\\
  \footnotetext[1]{e-mail: t96@nikhefh.nikhef.nl,$\;$ research supported by the Stichting FOM}
  \footnotetext[2]{e-mail: t30@nikhefh.nikhef.nl,$\;$ research supported by the Stichting FOM}
  NIKHEF-H, Amsterdam, The Netherlands}\\
\vspace{2\baselineskip}
{\bf Abstract}
\end{center}
We show how information on the uniformity properties of a
point set employed in numerical multidimensional integration
can be used to improve the error estimate over the usual
\mc\ one. We introduce a new measure of (non-)uniformity
for point sets, and derive explicit expressions for the various
entities that enter in such an improved error estimate.
The use of Feynman diagrams provides a transparent and straightforward 
way to compute this improved error estimate.
\\

\vspace{7.5cm}
\begin{center}
{\it Preprint submitted to Computer Physics Communications}
\end{center}

\newpage
\pagestyle{plain}
\setcounter{page}{1}
\setcounter{footnote}{0}

\section{Introduction}
In numerical integration, an object of paramount importance is the
integration error, that is, the difference between the actual (unknown)
value of the integral and the obtained numerical estimate. Since
exact knowledge of the error would imply exact knowledge of the
integral (and hence would obviate the need for a numerical estimate),
in practice one has to settle for an estimate of this
error, which must be shored up by plausibility arguments.
In deterministic integration, where the set $X_N$ of $N$ integration
points is determined beforehand, a deterministic error
bound (that is, a true upper limit on the error) can often be
obtained, provided the integrand falls in some known class of
functions with, for instance, given smoothness properties.
The integration rule ({\it i.e.\/} the point set $X_N$)
can then be optimized to guarantee a rapid decrease of the
error with increasing $N$ \cite{lattices}.

Unfortunately, in many practical problems the integrand is
not particularly smooth. This happens, for instance, in particle
phenomenology, where integrands may contain discontinuities
corresponding to experimental cuts in the admissible phase space.
In such cases, one usually relies on {\em \mc\/} integration,
where the point set $X_N$ is assumed to be chosen from the ensemble
of random point sets\footnote{In practice, the points are 
{\em pseudo-random\/} rather than truly random, being generated
by a deterministic algorithm: however, we shall assume that the
random number generator employed is sufficiently well behaved
to satisfactorily mimic a truly random sequence.}.
In the derivation of the error estimate, we use the fact that the point set $X_N$ is a `typical' member of the ensemble of all such point sets, and average over this ensemble.
The error estimate is, then, probabilistic rather than deterministic;
but with some care reliable estimates can be obtained, as
implied by the various laws of large numbers.
 The \mc\ method
can be applied to the very large class of square integrable
integrands; on the other hand, there is the drawback
that the error decreases asymptotically only
as $1/\sqrt{N}$. The existing variance-reducing techniques can at
most decrease the coefficient of the error estimate, and not
its behaviour with $N$, which is a direct consequence of
the random structure of the point set.\\

Recently, the approach known as {\em \qmc\/} has received
considerable attention. Here, one attempts to construct points
sets $X_N$ that are `more uniform' (in terms of a given
measure of uniformity) than truly random ones. The existence of
error bounds \`{a} la Koksma-Hlawka \cite{koksma}
underlies the belief
that such {\em \qran\/} point sets can, indeed, lead to an
error that decreases faster than $1/\sqrt{N}$: the Roth bound
\cite{roth}
suggests that a behaviour as $(\log N)^c/N$ may be possible, where
$c$ is some constant depending on the dimensionality of the integral.
Many \qran\ sequences with asymptotically small non-uniformity
have been proposed, such as 
van der Corput/Halton \cite{halton}, Faure \cite{faure}, Sobol'\cite{sobol}, 
and Niederreiter \cite{niederreiter}
sequences. These generally perform to satisfaction in many
applications: but, and this is the central problem
addressed in this and
subsequent papers, no reasonable way is yet known to estimate
the error while making use of the improved uniformity of the
point set. In fact, it is customary to just compute the error
estimate {\em as if\/} the point set $X_N$ were truly random.
This procedure tends to overestimate the error, as shown by
various case studies \cite{schlier}, 
so it is certainly safe to do so: but,
in our opinion, {\em \qmc\/ methods will come into their
own only when improved error estimates are available.\/} It is
our purpose to provide an attempt at such an improved
integration error estimate. To do this, we shall have to face the
fact that \qran\ point sets are {\em not\/} `typical'
members of an obvious ensemble: indeed, they are very special, with
a lot of ingenuity going in their construction. The central
idea of our approach is the definition of a suitable ensemble
of \qran\ point sets.\\

The lay-out of this paper is as follows. 
In section 2, we shall discuss
the error estimate for \mc, in such a way that it becomes clear
precisely which information on the \qran\ point set is needed to
improve the error estimate. To this end, we shall have to employ
{\em some\/} definition of (non-)uniformity, that is, a
{\em discrepancy.\/} In section 3, we shall introduce a particular
definition of a discrepancy, which we feel is better suited to
this kind of problem than the best-known discrepancy,
the so-called {\em star discrepancy\/} $D_N^\ast$, discussed
extensively in the literature \cite{kuipers}. 
In section 4, we shall discuss
in detail how the various ingredients in a new error estimate
can be computed, for a point set with a given, known, value of the
discrepancy. We shall do this by essentially combinatorial arguments,
using a technique based on Feynman diagrams. It will appear that,
in the limit of large $N$, our approach is closely related to
the computation of path integrals in field theory.

Finally, we have to apply in practice what we have
learned in theory. We shall
make such attempts, in the academic but instructive case of one-dimensional integral, and in more dimensions: these points will
be treated elsewhere \cite{nextpapers}.

\newpage
\section{Estimates, ensembles and discrepancy}
\subsection{The improved error estimate}
To set the stage, we define the $D$-dimensional integration region
$K$ to be $[0,1)^D$, the archetypical unit hypercube. 
The integrand is a function $f(x)$ defined on $K$, 
which may contain discontinuities but must be
at least square integrable. Its integral is denoted by $J_1$, where
\begin{equation}
J_p = \intk\;dx\;\left(f(x)\right)^p\;\;,\;\;p=1,2,\ldots\;\;.
\end{equation}
The point set $X_N$ consists of $N$ $D$-dimensional points
$x_k$, $k=1,2,\ldots,N$. Where necessary, we shall denote individual
components of points by Greek indices, $x_k\umu$, $\mu=1,2,\ldots,D$.
The numerical estimate of the integral is then given by
\begin{equation}
S = {1\over N}\suml_{k=1}^N f(x_k)\;\;,
\end{equation}
and the error made is then simply $\eta\equiv S-J_1$. The salient
fact about \mc\ is that the point set $X_N$ is a random object, and so,
therefore, is the error $\eta$. The standard \mc\ estimate is
derived by assuming that $X_N$ is a `typical' member of the ensemble
of point sets, governed by a probability density
$P_N(x_1,x_2,\ldots,x_N)$. We shall also define marginal densities:
\begin{equation}
P_k(x_1,x_2,\ldots,x_k) \equiv
 \intk\;dx_{k+1}dx_{k+2}\cdots dx_N\;
P_N(x_1,x_2,\ldots,x_N)\;\;,
\end{equation}
for $k=0,1,2,\ldots,N$, so that $P_0=1$. For truly random points, we have the ideal iid uniform distribution:
\begin{equation}
P_N(x_1,x_2,\ldots,x_N) = 1\;\;.
\end{equation}
We want, however, to be more general, and we shall write
\begin{equation}
P_k(x_1,x_2,\ldots,x_k) = 1 - {1\over N}F_k(x_1,x_2,\ldots,x_k)\;\;.
\label{definefk}
\end{equation}
Obviously, none of the $F_k$ can exceed $N$, 
and for truly random points they are identically zero. 
Moreover, since the order in which the points enter is immaterial 
in this integration problem, we shall assume that all $P_k$ 
and $F_k$ are invariant under any permutation of the arguments.

We are now ready to estimate the error, by computing the various
moments of the probability distribution of $\eta$. Denoting by brackets
the average over $P_N$, we have for the first moment
\begin{equation}
\langle\eta\rangle \;\;=\;\;
- J_1 + {1\over N}\suml_{k=1}^N \langle f(x_k)\rangle\nl
\;\;=\;\; {1\over N}\intk dx\;f(x)F_1(x)\;\;.
\end{equation}
We see that, if the integration is to be unbiased for all integrands,
we needs must have 
\begin{equation}
F_1(x)=0\;\;,
\label{f1zero}
\end{equation}
which means that, {\em individually,}
each integration point $x_k$ in $X_N$ must be uniformly distributed,
and the difference between \qran\ and truly random point sets
may show up only in correlations between points. Assuming this to be
indeed the case, we then turn to the second moment:
\begin{eqnarray}
\langle\eta^2\rangle & = & 
J_1^2 - {2J_1\over N}\suml_{k=1}^N\langle f(x_k)\rangle
+ {1\over N^2}\suml_{k,l=1}^N \langle f(x_k)f(x_l)\rangle\nl
& = & {1\over N}\left(J_2-J_1^2
-\left(1-{1\over N}\right)\;\intk dx_1dx_2\;f(x_1)f(x_2)F_2(x_1,x_2)
\right)\;.
\label{etasquared}
\end{eqnarray}
The following facts become apparent from this result. In the first
place, for truly random points, $F_2$ vanishes and we recover the
standard \mc\ estimate. Secondly, only a small, $\order{1/N}$, deviation
in $P_N$ from the truly random uniform iid case
can already significantly diminish the error,
due to the delicate cancellations involved in the computation
of $\langle\eta^2\rangle$. Finally, the mechanics behind this
improvement become obvious: in very uniform point sets
(such as \qran\ point sets), that have a low discrepancy, the 
points are spread `more evenly' than in typical random sets: they
`repel' each other, leading to a positive value of $F_2(x_1,x_2)$
whenever $x_1$ and $x_2$ are `close' in $K$: this suppresses the
contribution to $\langle\eta^2\rangle$ from regions where
$f(x)^2$ tends to be large. Conversely, for very non-uniform sets, where
the points are more `clumped' together, and whose
discrepancy is large, $F_2$ will be negative
for neighbouring points, with a corresponding punishment in the
error estimate. A very simple and instructive illustration
of this idea is given in Appendix B.

A final remark is in order here: for truly
random  point sets, the usefulness of the error estimate relies
on the fact that the distribution of $\eta$ tends to a Gaussian,
as implied by the Central Limit Theorem. In principle, we ought
to reconstruct a proof of this theorem for the more general $P_N$
discussed here. Although we have not yet done so, we have good 
reason to believe that a Central Limit Theorem holds
in our case as well, with an implied Gaussian width following
from \eqn{etasquared}.

\subsection{Generating functions}
Our task is now clear: we have to find, for \qran\ point sets,
a workable definition of $P_N$, and a corresponding formula for
$F_2$. To do so, we first assume that there exists {\em some\/}
measure of (non-)uniformity of point sets $X_N$: that is, there must
be given some {\em discrepancy\/} $D_N$ 
as a function of the $x_k$, and we shall also assume that we can, 
for a given $X_N$, compute its value which we shall denote by $s$:
\begin{equation}
D_N(x_1,x_2,\ldots,x_N) = s\;\;.
\end{equation}
A `small' value of $s$ shall indicate a point set that is relatively
uniform compared to a truly random one. We shall defer an explicit 
definition of $D_N$ to the next section, and for the moment just 
assume that one is given. We then propose to use for $P_N$ the
probability density obtained by assuming that all points are uniformly
distributed, with the additional constraint that the discrepancy
takes on the value $s$:
\begin{eqnarray}
P_N(s;x_1,\ldots,x_N) & \equiv & 
{H_N(s;x_1,\ldots,x_N)\over H_0(s)}\;\;,\nl
H_k(s;x_1,\ldots,x_k) & \equiv & 
\intk dx_{k+1}\cdots dx_N\;\delta(D_N(x_1,\ldots,x_N)-s)\;\;,
\end{eqnarray}
where we have introduced the Dirac delta distribution to handle
the discrepancy constraint. The function $H_0(s)$ is, then,
the probability distribution of $s$ over the ensemble of
all truly random point sets, which is an interesting quantity
in its own right. The idea behind this definition is the following.
We are not allowed, in principle, to consider a \qran\ point set 
as a `typical' one in the whole random ensemble, since $s$ is
(hopefully) small compared to the expected value for random sets:
but, in the subset with that particular value of $s$, it may very
well be typical (indeed, this is the same fingers-crossed attitude
that allows us to use, for pseudo-random point sets, the standard \mc\ error estimate in the first place). 
Moreover, if no information whatsoever is available
on the value of $s$, we have to integrate over all $s$ with probability
density $H_0(s)$, upon which the delta function constraint drops out and we are back in the truly random case.\\

We now proceed to calculate $F_2$. Let us define a set of 
moment-generating functions as follows:
\begin{equation}
G_k(z;x_1,\ldots,x_k) \equiv\intk dx_{k+1}\cdots dx_N\;
\exp\left(zD_N(x_1,\ldots,x_N)\right)\;\;,
\end{equation}
that is, the generating function where the first $k$ of the integration points are kept fixed, and the remaining ones are 
integrated over: $G_0(z)$, then, is the moment-generating
function for $H_0(s)$. We employ the definition of the Dirac delta
distribution as a Fourier/Laplace transform to write
\begin{equation}
H_k(s;x_1,\ldots,x_k) = {1\over2\pi i}\intinfi
dz\;e^{-zs}G_k(z;x_1,\ldots,x_k)\;\;,
\end{equation}
where the integration contour must run to the left of any
singularities. So,
knowledge of the $G_k$ allows us to compute everything we need.
We may uniquely split each $G_k$ into a constant part, and
a part that depends on the $x_{1,\ldots,k}$ and averages
out to zero:
\begin{eqnarray}
& & {G_k(z;x_1,\ldots,x_k)\;\; =\;\;  
G_0(z) + {1\over N}\rho_k(z;x_1,\ldots,x_k)\;\;,}\nl
& &\intk dx_1\cdots dx_k\;\rho_k(z;x_1,\ldots,x_k)\;\; =\;\;  0\;\;.
\end{eqnarray}
It follows that we can express $F_k$ as
\begin{eqnarray}
F_k(s;x_1,\ldots,x_k) & = & -{R_k(s;x_1,\ldots,x_k)/H_0(s)}\;\;,\nl
R_k(s;x_1,\ldots,x_k) & = & {1\over2\pi i}\intinfi
dz\;e^{-zs}\rho_k(z;x_1,\ldots,x_k)\;\;.
\end{eqnarray}
In this way, we can compute the necessary two-point correlation
$F_2$, provided we can compute, at least in the form of an
asymptotic expansion in $1/N$, the moment-generating functions $G_k$:
the dominant contribution to $F_k$ will then come from the
leading terms in $G_0$ and $\rho_k$.
This depends, of course, on a manageable choice of the discrepancy
$D_N$, and this will be the subject of the next section.
We wish to point out that, actually, an exact expression for the
$F_k$ would allow us to determine the lowest possible value for $s$,
simply by putting $F_k(s;x_1,\ldots,x_k)=N$; from the fact that it is
so hard to improve on the Roth bound, we may infer that an asymptotic
expansion in $1/N$ is probably the best possible result we can hope
for at this moment.
Finally, we remark that a check on our results for $R_2(s;x_1,x_2)$
is provided by the identities
\begin{equation}
\intk dx_2\;R_2(s;x_1,x_2) = 0\;\;\;,\;\;\;
\intl_0^{\infty} ds\;R_2(s;x_1,x_2) = 0\;\;,
\end{equation}
where the first identity follows from \eqn{f1zero}, and the
second one by construction.

\section{The Fourier discrepancy}
We now turn to a useful definition of a discrepancy measure for $X_N$.
This we do by first constructing a model of the class of integrands
likely to be encountered, and then considering the squared integration
error expected for this class of integrands. This procedure is 
analogous to that employed by Wo\'{z}niakowski \cite{wozni}.\\

\subsection{The one-dimensional case}
For simplicity, we shall first discuss the case $D=1$, and only further
on generalize to more dimensions. We start by assuming that our
integrands admit of a decomposition into a set of
orthonormal functions. These satisfy
\begin{equation}
\intk dx\;u_m(x)u_n(x) = \delta_{mn}\;\;\;,\;\;\;
\suml_{n\ge0}u_n(x)u_n(y) = \delta(x-y)\;\;.
\end{equation}
A general integrand $f(x)$ can then be written as
\begin{equation}
f(x) = \suml_{n\ge0}\;v_n\;u_n(x)\;\;,
\end{equation}
so that the coefficients $v_n$ determine the function. Since
these coefficients form an enumerable set, it is fairly easy to set up a combined probability distribution for them, which gives us then a
probability measure $\ddf$ on the class of integrands. We take the
following choice:
\begin{equation}
\ddf  = \prol_{n\ge0}\;dv_n\;{\exp(-v_n^2/2\si_n^2)\over\sqrt{2\pi\si_n^2}}\;\;,
\end{equation}
so that each coefficient $v_n$ is distributed normally with 
mean value 0 and standard deviation $\si_n$. We shall call $\si_n$
the {\em strength\/} of the mode $u_n(x)$ in our function ensemble.
Denoting by $\langle\rangle_f$ an average over the ensemble under the
measure $\ddf$, we then have
\begin{equation}
\langle v_n\rangle_f = 0\;\;\;,\;\;\;
\langle v_nv_m\rangle_f = \si_m^2\delta_{mn}\;\;.
\end{equation}
We can also compute the moments of the integration error $\eta$
over the integrand ensemble. The integral $J_1$ is of course just
given by $v_0$, and moreover we have
\begin{equation}
\langle\eta\rangle_f = 0\;\;\;,\;\;\;
\langle\eta^2\rangle_f =
{1\over N^2}\suml_{k,l=1}^N \beta_1(x_k,x_l)\;\;,
\end{equation}
where we have introduced functions $\beta_m$ as
\begin{equation}
\beta_m(x,y) = \suml_{n>0}\;\si_n^{2m} u_n(x)u_n(y)\;\;.
\end{equation}
These functions, which will play an important r\^{o}le in our discussion,
are symmetric in their arguments, and have the properties that
\begin{equation}
\intk dy\;\beta_m(x,y) = 0\;\;\;,\;\;\;
\intk dy\;\beta_m(x_1,y)\beta_n(x_2,y) = \beta_{m+n}(x_1,x_2)\;\;,
\label{betaproperties}
\end{equation}
for $m,n>0$.
In fact it is also simple to prove that the error $\eta$ has
a Gaussian distribution, not as a consequence of any law of large
numbers, but because of the Gaussian character of our measure
$\ddf$ on the space of integrands.\\

Up to this point, we have not had to specify the particular
orthonormal functions. In practice, we shall use the Fourier basis:
\begin{equation}
u_0(x) = 1\;\;,\;\;
u_{2n-1}(x) = \sqrt{2}\;\sin(2\pi nx)\;\;,\;\;
u_{2n}(x) = \sqrt{2}\;\cos(2\pi nx)\;,
\end{equation}
where $n$ runs over the positive integers.

In addition, we shall make one very important assumption,
namely that the sines and cosines have equal strength:
\begin{equation}
\si_{2n-1} \equiv \si_{2n}\;\;.
\end{equation}
We shall call this property {\em translational invariance}.
Its physically reasonable motivation is the fact that under this assumption the mode with frequency $n$ (made up from $u_{2n-1}$ and
$u_{2n}$) has a uniformly distributed phase. Translational invariance will enable us to considerably simplify our results. Most
importantly, it leads us to write
\begin{equation}
\beta_m(x_1,x_2) = \suml_{n>0}2(\si_{2n})^{2m}
\cos(2\pi n(x_1-x_2))\;\;,
\end{equation}
so that the functions $\beta$ only depend on the {\em difference\/}
between $x_1$ and $x_2$, and we may also write 
$\beta_m(x_1-x_2)$. Another attractive consequence is the fact that
(as we shall see) point sets that only differ by a translation 
(assuming a periodic extension of $K$ and $X_N$)
have the same discrepancy, in contrast to, {\it e.g.}, the star discrepancy $D_N^\ast$.\\

The above consideration leads us to propose, as an appropriate
definition of discrepancy, the following:
\begin{equation}
D_N = D_N(x_1,\ldots,x_N) \equiv 
{1\over N}\suml_{k,l=1}^N\;\beta_1(x_k,x_l)\;\;.
\end{equation}
Note that we have taken out one factor of $1/N$: this will 
make the discrepancy independent of $N$ for truly random points.
It is easily seen that, for truly random points,
\begin{equation}
\langle D_N\rangle = \suml_{n>0}\si_n^2\;\;.
\end{equation}

There are a few additional observations in order here. In the first 
place, note that the r\^{o}les of integrand and point set have,
in some sense, been reversed here: in the derivation of the
standard \mc\ error, we keep the integrand fixed, and average over
the ensemble of point sets, upon which $\langle\eta^2\rangle$
no longer depends on the particular point set but only upon a
property of the integrand (namely its variance); while here we
have kept the point set fixed, and averaged over an ensemble of integrands, so that  $\langle\eta^2\rangle_f$ no longer depends
on the integrand, but only upon a property of the point set $X_N$,
namely its discrepancy. A number of other results along these lines
have been presented in \cite{cpccomplexity}. In the second place,
the zero mode with $n=0$ does not enter in the discrepancy: this is
reasonable since $v_0$ is in fact the desired integral, and the
{\em error\/} just consists in our partial inability to get rid of the higher modes, with $n>0$. Finally, the strengths $\si_n$ cannot
be chosen too arbitrarily. If our average integrand is to be
quadratically integrable, we need to have
\begin{equation}
\left\langle\intk dx\;f(x)^2\right\rangle_{\!\!\!f}
= \suml_{n\ge0}\sigma_n^2 < \infty\;\;\;,
\end{equation}
while additional smoothness requirements will ask for an 
even faster convergence of this sum. An admissible, and reasonable,
choice for the $\si_n$ will be, for instance
\begin{equation}
\si_0 = 1\;\;,\;\;\si_{2n} = 1/n\;\;,\;\;n>0\;\;,
\end{equation}
and this we shall consider in the practical applications \cite{nextpapers}.

\subsection{The more-dimensional case}
The extension of the above discussion to more dimensions is
quite straightforward. 
We shall use again $x$ to denote $D$-dimensional points.
The complete set of orthonormal
functions is now enumerated not by a scalar $n$, but by a vector
quantity $\vec{n}$ with indices $n\umu\ge0$, with $\mu=1,2,\ldots,D$.
The set of orthonormal functions is now given by
\begin{equation}
u_{\vec{n}}(x) \equiv \prol_{\mu=1}^Du_{n\umu}(x\umu)\;\;;
\end{equation}
the measure $\ddf$ is straightforward, with the strengths of the
various modes denoted by $\si_{\vec{n}}$, and the discrepancy
is again given by
\begin{equation}
D_N = D_N(x_1,\ldots,x_N) \equiv
{1\over N}\suml_{k,l=1}^N\;\beta_1(x_k,x_l)\;\;,
\label{discrepancydefinition}
\end{equation}
where
\begin{equation}
\beta_m(x,y) = \suml_{\vec{n}>0}\si_{\vec{n}}^{2m}u_{\vec{n}}(x)u_{\vec{n}}(y)\;\;.
\end{equation}
Here, the sum runs over all $\vec{n}$ except $\vec{n}=(0,0,\ldots,0)$.
Again, for truly random points we have
\begin{equation}
\langle D_N\rangle = \suml_{\vec{n}>0}\si_{\vec{n}}^2\;\;.
\end{equation}
Translational invariance in $D$ dimensions requires
\begin{eqnarray}
\si_{(2n^1,2n^2,\ldots,2n^D)} & = &  
\si_{(2n^1-1,2n^2,\ldots,2n^D)} = \cdots\nl
\cdots & = & \si_{(2n^1,2n^2,\ldots,2n^D-1)} = \cdots\nl
\cdots & = & \si_{(2n^1-1,2n^2-1,\ldots,2n^D-1)}\;\;,
\end{eqnarray}
so that the strengths must be equal in groups of $2^{D-k}$, 
where $k$ is the number of vanishing components of $\vec{n}$. 
Again, for our integrands to be quadratically integrable on the average,
we must have
\begin{equation}
\suml_{\vec{n}>0}\si_{\vec{n}}^2 < \infty\;\;\;.
\end{equation}
Now, we may choose to let the strengths be dominated by the
highest partial frequency, for instance
\begin{equation}
\si_{\vec{n}} = \si(\max_{\mu}n\umu)\;\;.
\end{equation}
Because of the increasing multiplicity with
$\max_{\mu}n\umu$, $\si_{\vec{n}}$ must then decrease more
rapidly than in the one-dimensional case. Other reasonable
choices, such as $\si_{\vec{n}}=\si(\sum_{\mu}n\umu)$, lead to
the same conclusion. 
As usual, we encounter here the phenomenon that a smooth function
in one dimension is not precisely the same concept as the
projection onto one dimension of a function that is smooth in
many dimensions: the `curse of dimensionality' crops up again,
in disguise.\\

We have now finished the first part of our program, that is, the
establishing of a reasonable definition of a discrepancy. We
want to remark that it differs from the better known 
{\em star-discrepancy\/} $D_N^{\ast}$,
discussed so extensively in \cite{kuipers}. That one can, in fact,
also be derived from a class of integrands, as first made explicit
in \cite{wozni}. The appropriate integrand class is defined by taking
for $\ddf$ the Wiener sheet measure; although mathematically
attractive, this is by no means the preferred choice for many
practical applications, since it singles out functions that are
everywhere continuous but nowhere differentiable. Moreover,
the translational invariance will lead, as we shall see,
to $F_1(s;x_1)\equiv0$, and the lack of this invariance for the
star-discrepancy probably implies that we cannot simply
prove that it leads to unbiased integral estimates. We should like
to point out that, in fact, we have been able to derive the form of $H_0(s)$ for the star-discrepancy as well \cite{fred}.

\section{Discrepancies by Feynman diagrams}
We must now proceed with our program, and find expressions for 
\begin{eqnarray}
\lefteqn{G_k(z;x_1,\ldots,x_k) = \suml_{m\ge0}{z^m\over m!}
\langle D_N^m\rangle_k\;\;,}\nl
\lefteqn{\langle D_N^m\rangle_k =}\nl
& & {1\over N^m}
\intk dx_{k+1}\cdots dx_N
\suml_{r_{1,\ldots,2m}=1}^N
\beta_1(x_{r_1},x_{r_2})\cdots\beta_1(x_{r_{2m-1}},x_{r_{2m}})\;\;.
\end{eqnarray}
We shall call the points $x_1,\ldots,x_k$, that are kept fixed,
{\em external points}, and the remaining $N-k$ ones, that
are integrated over, {\em internal points}.
Note that, in the multiple sum, each index $r_j$ runs
over both external and internal points. The various internal
points are essentially indistinguishable, and they give rise to
combinatorial factors; the study of these factors will enable us to
write $\langle D_N^m\rangle_k$ as an asymptotic series in $1/N$.

The part of the multiple sum where all indices $r_1,\ldots,r_{2m}$ are equal and internal gives rise to a single combinatorial factor $N-k$;
if they are distributed over two different internal values, we shall
have a combinatorial factor $(N-k)(N-k-1)$, and so on. It follows that
the largest combinatorial factor that can, in principle, occur, is
the one for the case where all indices are internal and distinct.
However, that contribution consists of separate factors
$$
\intk dx_{r_i}dx_{r_j}\beta_1(x_{r_i},x_{r_j})\;\;,
$$
which vanish. Therefore, the {\em actually\/} largest combinatorial
factor is that for the situation where the indices are all internal,
and fall in distinct pairs: the corresponding combinatorial factor is
\begin{eqnarray}
(N-k)\fal{m} & = & (N-k)!/(N-k-m)!\nl
& = & N^m - N^{m-1}\left[{m\fal{2}\over2} + km\right]
+ \order{N^{m-2}}\;\;.
\end{eqnarray}
The `falling power' notation is taken from Graham {\it et al.\/}
\cite{graham};
its asymptotic expansion for large $N$ is derived in  Appendix A.

\subsection{Feynman diagrams}
We are now ready to compute the first few terms in the $1/N$ expansion
of $\langle D_N^m\rangle_k$. We start by introducing a diagrammatical
technique. Each internal point we shall denote by a dot, and each
external point by a cross. Every function $\beta_1$ shall be denoted
by a {\em link\/}, a solid line between dots or crosses. 
Every contribution to $\langle D_N^m\rangle_k$ will therefore
contain precisely $m$ links. Identical values of the
summation indices mean that the corresponding points will be
contracted, leading to {\em vertices\/} with two, three, four, \ldots legs. Neglecting for the moment the combinatorials we have,
for instance,
\begin{eqnarray}
\bpic(30,4)(0,0)\Cross{0}{0}{4}{4}\Line(2,2)(28,2)\Cross{26}{0}{30}{4}
\Text(2,-5)[c]{$x_1$}\Text(28,-5)[c]{$x_2$}\epic 
& = & \beta_1(x_1,x_2) \;\;,\nl
\bpic(30,4)(0,0)\Cross{0}{0}{4}{4}\Line(2,2)(28,2)\Vix{28,2}
\Text(2,-5)[c]{$x$}\epic
& = & \intk dy\;\beta_1(x,y) = 0\;\;,\nl
\bpic(30,4)(0,0)\Vix{2,2}\Line(2,2)(28,2)\Vix{28,2}\epic 
& = &\intk dy_1dy_2\;\beta_1(y_1,y_2) = 0 \;\;,\nl
\bpic(20,20)(0,7)\BC(10,10){10}\Vix{10,20}\epic
& = & \intk dy\;\beta_1(y,y)\;\;,\nl
\bpic(60,20)(0,7)\BC(10,10){10}\Line(20,10)(40,10)
\BC(50,10){10}\Vix{20,10}\Vix{40,10}\epic
& = & \intk dy_1dy_2\;
\beta_1(y_1,y_1)\beta_1(y_1,y_2)\beta_1(y_2,y_2)\;\;, 
\label{diagexamples}
\end{eqnarray}
where the two zero results follow from \eqn{betaproperties}.
In each diagram or product of diagrams, the
combinatorial factor can be read off immediately from the
number of internal points in evidence: 
if there are $p$ internal points, the factor is
$(N-k)\fal{p}$. External points do not contribute a combinatorial
factor, but it must be kept in mind that we shall have to sum
over all external points. For instance, the diagram
$\bpic(20,4)(0,0)\Cross{0}{0}{4}{4}\Line(2,2)(18,2)
\Cross{16}{0}{20}{4}\epic$ must be interpreted as
\begin{equation}
\bpic(20,4)(0,0)\Cross{0}{0}{4}{4}\Line(2,2)(18,2)
\Cross{16}{0}{20}{4}\epic\;\; =\;\; 
\suml_{\stackrel{k,l=1}{k\ne l}}^N\;\;
\bpic(30,4)(0,0)\Cross{0}{0}{4}{4}\Line(2,2)(28,2)
\Cross{26}{0}{30}{4}
\Text(5,-8)[c]{$x_k$}\Text(28,-8)[c]{$x_l$}\epic\;\;,
\end{equation}
so that this diagram actually contains $k(k-1)$ terms.

In writing out the $1/N$ expansion, we can simply determine
which kinds of diagrams will contribute, as follows. The leading
power of a two-point internal vertex is 1, that of a three-point 
internal vertex is $1/\sqrt{N}$, for a four-point internal
vertex we have $1/N$, and so on. 
Formally, this is similar to a
field theory with a universal coupling constant $g$ proportional to
$1/\sqrt{N}$. Moreover, each external point will effectively carry
a factor $1/\sqrt{N}$. 
If we decide to keep only the terms up to and including $1/N$, we
shall have to allow for at most two external points,
or a single external point and one internal three-vertex, or two
internal three-vertices, or a single internal four-vertex.
In a field theory, this corresponds to the first-order
correction, proportional to $g^2$.
Moreover, our property of translational invariance implies
precisely the analogue for a field theory, namely momentum
conservation\footnote{The analogy with a real field theory cannot
be carried too far, however, since such a theory has an infinite
number of degrees of freedom. In the present case, all vertices
and combinatorial factors also carry subleading contributions,
which must be properly taken into account.}. Also, the external
points, that act somewhat like sources, need rescaling
by factors $\sqrt{N}$, which could be envisioned as the truncation
process by which a Green's function is turned into an $S$-matrix element. This is precisely what the extraction of the factor $1/N$
in \eqn{definefk} does for us, and so we recognize what the
function $F_2(s;x_1,x_2)$ actually is: it is the full
propagator.\\

\subsection{An illustration: the first two moments}
We shall now show how to apply the diagrammatic techniques in
the calculation of $G_k(z;x_1,\ldots,x_k)$.
The first nontrivial term comes from $\langle D_N\rangle_k$,
which we can write diagrammatically as
\begin{eqnarray}
\langle D_N\rangle_k & = & {1\over N}\left[\;\;
\bpic(20,4)(0,0)\Cross{0}{0}{4}{4}\Line(2,2)(18,2)
\Cross{16}{0}{20}{4}\epic
+ \bpic(20,20)(0,7)\BC(10,10){10}\Cross{8}{18}{12}{22}\epic
+ 2(N-k)\;\bpic(20,4)(0,0)\Cross{0}{0}{4}{4}\Line(2,2)(18,2)
\Vix{18,2}\epic\right.\nl
& & \hphantom{{1\over N}}\left.
+ (N-k)\;\bpic(20,20)(0,7)\BC(10,10){10}\Vix{10,20}\epic
+ (N-k)\fal{2}\;\bpic(20,2)(0,0)\Vix{2,2}\Line(2,2)(18,2)\Vix{18,2}\epic
\;\;\right]\;\;.
\end{eqnarray}
We may considerably simplify this. In the first place, we have
$\bpic(20,2)(0,0)\Cross{0}{0}{4}{4}\Line(2,2)(18,2)\Vix{18,2}\epic = 0$
and $\bpic(20,2)(0,0)\Vix{2,2}\Line(2,2)(18,2)\Vix{18,2}\epic = 0$,
see \eqn{diagexamples}. In addition, the property of translational
invariance has the important consequence that a diagram with only
a single external point evaluates to a constant. For instance,
\begin{equation}
\bpic(20,20)(0,7)\BC(10,10){10}\Cross{-2}{8}{2}{12}
\Text(-6,10)[c]{$x$}\epic\;\;
= \beta_1(x,x) = \beta_1(0) = \intk dy\;\beta_1(y,y)
=\;\; \bpic(20,20)(0,7)\BC(10,10){10}\Vix{0,10}\epic\;\;.
\end{equation}
This also implies that, whenever two parts of a diagram are
connected by a single vertex, we may split it up (of course, 
without changing the combinatorial factor!), so that, for
instance,
\begin{eqnarray}
\bpic(40,10)(0,7)\BC(10,10){10}\BC(30,10){10}\Vix{20,10}\epic & = &  
\intk dx\;\beta_1(x,x)^2 = \left(\intk dx\;\beta_1(x,x)\right)^2 = 
\left(\bpic(20,20)(0,7)\BC(10,10){10}\Vix{10,20}\epic\right)^2\;\;,\nl 
\bpic(60,20)(0,7)\BC(10,10){10}\Line(20,10)(40,10)\BC(50,10){10}
\Vix{20,10}\Vix{40,10}\epic & = & 
\left(\bpic(20,20)(0,7)\BC(10,10){10}\Vix{10,20}\epic\right)^2\;
\bpic(20,4)\Vix{0,2}\Vix{20,2}\Line(0,2)(20,2)\epic \;\;= 0\;\;;
\end{eqnarray}
the last line is an example of the more general phenomenon that
all {\em tadpole\/} diagrams with only internal points 
evaluate to zero. Again, this is due to
our assumption of translational invariance: incidentally, it
immediately proves that 
\begin{equation}
G_1(z;x_1) = G_0(z)\;\;\;\mbox{to all orders in $1/N$,}
\end{equation}
which in its turn implies that
\begin{equation}
F_1(s;x_1) = 0\;\;\;\mbox{to all orders in $1/N$,}
\end{equation}
as required in \eqn{f1zero}. We can now write $\langle D_N\rangle_k$ as
\begin{equation}
\langle D_N\rangle_k\;\;=\;\; 
\bpic(20,20)(0,7)\BC(10,10){10}\Vix{10,20}\epic
\;\;+\;\;
{1\over N}\;\bpic(20,4)(0,0)\Cross{0}{0}{4}{4}\Line(2,2)(18,2)
\Cross{16}{0}{20}{4}\epic\;\;,
\end{equation}
where we must keep in mind that the second term implies a summation
over the $k(k-1)$ pairs of {\em different\/} points in $x_1,x_2,\ldots,x_k$. It follows trivially that, as required,
\begin{eqnarray}
\lefteqn{\intk dx_k\;\langle D_N\rangle_k = \langle D_N\rangle_{k-1}
\;\;\Rightarrow}& &\nl
& \Rightarrow &
\intk dx_k\;F_k(s;x_1,\ldots,x_k) = 
F_{k-1}(s;x_1,\ldots,x_{k-1})\;\;.
\end{eqnarray}
We now proceed to the next order, and
compute $\langle D_N^2\rangle_k$. The only contributions
that do not immediately vanish under \eqn{betaproperties} are
\begin{eqnarray}
\langle D_N^2\rangle_k & = &
{1\over N^2}\left[
\bpic(20,20)(0,7)\Cross{0}{0}{4}{4}\Line(2,2)(18,2)\Cross{16}{0}{20}{4}
\Cross{0}{16}{4}{20}\Line(2,18)(18,18)\Cross{16}{16}{20}{20}\epic
\;\;+\;\; 
4(N-k)\;\bpic(30,4)\Cross{0}{0}{4}{4}\Line(2,2)(28,2)\Vix{15,2}
\Cross{26}{0}{30}{4}\epic
\;\;+\;\; 
4\;\bpic(30,4)\Cross{0}{0}{4}{4}\Line(2,2)(28,2)\Cross{13}{0}{17}{4}
\Cross{26}{0}{30}{4}\epic\right.\nl
& &\;\;+\;\; 
2(N-k)\;\bpic(30,20)(0,11)\BC(10,20){10}\Vix{10,10}
\Cross{0}{0}{4}{4}\Line(2,2)(18,2)\Cross{16}{0}{20}{4}\epic
\;\;+\;\; 
2\;\bpic(30,20)(0,10)\BC(10,20){10}\Cross{8}{8}{12}{12}
\Cross{0}{0}{4}{4}\Line(2,2)(18,2)\Cross{16}{0}{20}{4}\epic
\;\;+\;\;
4\;\bpic(35,20)(0,7)\BC(10,10){10}\Cross{18}{8}{22}{12}
\Line(20,10)(33,10)\Cross{31}{8}{35}{12}\epic \nl
& & \;\;+\;\;
2\;\bpic(20,20)(0,7)\BC(10,10){10}\Cross{-2}{8}{2}{12}
\Cross{18}{8}{22}{12}\epic
\;\;+\;\;
4(N-k)\;\bpic(20,20)(0,7)\BC(10,10){10}\Cross{-2}{8}{2}{12}
\Vix{20,10}\epic
\;\;+\;\;
2(N-k)\fal{2}\;\bpic(20,20)(0,7)\BC(10,10){10}
\Vix{0,10}\Vix{20,10}\epic\nl
& &\;\;+\;\;
\bpic(45,20)(0,7)\BC(10,10){10}\Cross{8}{-2}{12}{2}
\BC(35,10){10}\Cross{33}{-2}{38}{2}\epic
\;\;+\;\;
2(N-k)\;\bpic(45,20)(0,7)\BC(10,10){10}\Cross{8}{-2}{12}{2}
\BC(35,10){10}\Vix{35,0}\epic\nl
& &\;\;+\;\;
(N-k)\fal{2}\;\bpic(45,20)(0,7)\BC(10,10){10}\Vix{10,0}
\BC(35,10){10}\Vix{35,0}\epic\nl
& &\;\;+\;\;\left.
\bpic(40,20)(0,7)\BC(10,10){10}\BC(30,10){10}
\Cross{18}{8}{22}{12}\epic 
\;\;+\;\;
(N-k)\;\bpic(40,10)(0,7)\BC(10,10){10}\BC(30,10){10}\Vix{20,10}\epic\;\;
\right]\;\;.
\end{eqnarray}
Using translational invariance, we may rewrite this as
\begin{eqnarray}
\langle D_N^2\rangle_k & = & {1\over N^2}\left[
\bpic(20,20)(0,7)\Cross{0}{0}{4}{4}\Line(2,2)(18,2)\Cross{16}{0}{20}{4}
\Cross{0}{16}{4}{20}\Line(2,18)(18,18)\Cross{16}{16}{20}{20}\epic
\;\;+\;\; 
4(N-k)\;\bpic(30,4)\Cross{0}{0}{4}{4}\Line(2,2)(28,2)\Vix{15,2}
\Cross{26}{0}{30}{4}\epic
\;\;+\;\; 
4\;\bpic(30,4)\Cross{0}{0}{4}{4}\Line(2,2)(28,2)\Cross{13}{0}{17}{4}
\Cross{26}{0}{30}{4}\epic\right.\nl
& & \;\;+\;\; 
2N\;\bpic(30,20)(0,11)\BC(10,20){10}\Vix{10,10}
\Cross{0}{0}{4}{4}\Line(2,2)(18,2)\Cross{16}{0}{20}{4}\epic
\;\;+\;\;
2(N\fal{2}-k\fal{2})\;\bpic(20,20)(0,7)\BC(10,10){10}
\Vix{0,10}\Vix{20,10}\epic\nl
& &\;\;+\;\;\left.
2\;\bpic(20,20)(0,7)\BC(10,10){10}\Cross{-2}{8}{2}{12}
\Cross{18}{8}{22}{12}\epic
\;\;+\;\;
N^2\!\!\left(\bpic(20,20)(0,7)\BC(10,10){10}\Vix{10,20}\epic\right)^2\;\;
\right]\;\;.
\end{eqnarray}
It can again easily be checked that
\begin{equation}
\intk dx_k\;\langle D_N^2\rangle_k = 
\langle D_N^2\rangle_{k-1}\;\;.
\end{equation}
If we restrict ourselves to terms of order $\order{1}$ and $\order{N^{-1}}$, we have
\begin{eqnarray}
\langle D_N^2\rangle_k & = & 
\left(\bpic(20,20)(0,7)\BC(10,10){10}\Vix{10,20}\epic\right)^2
\;\;+\;\;{2\over N}\;
\bpic(20,30)(0,10)\BC(10,20){10}\Vix{10,10}
\Cross{0}{0}{4}{4}\Line(2,2)(18,2)\Cross{16}{0}{20}{4}\epic
\;\;+\;\;{2N\fal{2}\over N^2}\;
\bpic(20,20)(0,7)\BC(10,10){10}\Vix{0,10}\Vix{20,10}\epic
\;\;+\;\;{4\over N}\;\bpic(30,4)(0,-2)\Cross{0}{0}{4}{4}
\Line(2,2)(28,2)\Vix{15,2}\Cross{26}{0}{30}{4}\epic\;\;.
\end{eqnarray}

\subsection{The general result to leading order}
We shall now discuss the computation of all diagrams of
the leading necessary order. To start,
let us concentrate on the leading terms, of order $\order{1}$.
These are contributed by graphs containing only internal two-vertices,
that is, they must be of the form of a product of closed loops:
\begin{eqnarray}
G_k(z) & = &
W_0(m) + \order{{1\over N}}\;\;,\nl
W_0(m) & \equiv & \suml_{p_{1,2,3,\ldots}\ge0}
A(m;p_1,p_2,p_3,\ldots)
\left(\bpic(20,10)(0,7)\BC(10,10){10}
\Vix{10,20}
\epic\right)^{p_1}
\left(\bpic(20,10)(0,7)\BC(10,10){10}
\Vix{0,10}\Vix{20,10}
\epic\right)^{p_2}
\left(\bpic(20,10)(0,7)\BC(10,10){10}
\Vix{10,20}\Vix{3.5,3.5}\Vix{16.5,3.5}
\epic\right)^{p_3}
\cdots,
\end{eqnarray}
where we have left out the factor $N\fal{m}/N^m$, and imply the
constraint
\begin{equation}
m = p_1 + 2p_2 + 3p_3 + \cdots\;\;.
\label{pconstraint}
\end{equation}
The factor $A(m;p_1,p_2,\ldots)$ is governed by a recursion relation.
We may go from $m-1$ to $m$ by either adding a single one-link loop,
or by putting an extra link in any $k$-link loop, thereby
turning it into a $(k+1)$-link loop. The recursion relation 
therefore reads
\begin{eqnarray}
\lefteqn{A(m;p_1,p_2,p_3,\ldots) = 
A(m-1;p_1-1,p_2,p_3,\ldots)\;\; + }\nl
& &\suml_{k>0} 2k(p_k+1)
A(m-1;p_1,p_2,\ldots,p_k+1,p_{k+1}-1,\ldots) \;\;.
\label{arecursion}
\end{eqnarray}
We may use the Ansatz
\begin{equation}
A(m;p_1,p_2,p_3,\ldots) = 
{m!\over p_1!p_2!p_3!\cdots}Ca_1^{p_1}a_2^{p_2}a_3^{p_3}\cdots\;\;,
\end{equation}
with $C$ and the $a$'s to be determined. Putting this Ansatz in
\eqn{arecursion} leads us to
\begin{equation} m  = {p_1\over a_1} + \suml_{k>0} 2k{a_k\over a_{k+1}}p_{k+1}\;\;.
\end{equation}
Since the coefficients of the $p$'s are known from
\eqn{pconstraint}, this gives us the symmetry factor
associated with each closed $k$-link loop:
\begin{equation}
a_k = {2^k\over2k}\;\;.
\end{equation}
By inspection of $\langle D_N\rangle_k$, we also find $C=1$.
The result for the terms that give the leading contribution is therefore
\begin{equation}
W_0(m) \equiv \suml_{p_{1,2,3,\ldots}\ge0}
{m!\over p_1!p_2!p_3!\cdots}
\left({2^1\over2}\;\bpic(20,10)(0,7)\BC(10,10){10}
\Vix{10,20}
\epic\right)^{p_1}
\left({2^2\over4}\;\bpic(20,10)(0,7)\BC(10,10){10}
\Vix{0,10}\Vix{20,10}
\epic\right)^{p_2}
\left({2^3\over6}\;\bpic(20,10)(0,7)\BC(10,10){10}
\Vix{10,20}\Vix{3.5,3.5}\Vix{16.5,3.5}
\epic\right)^{p_3}
\cdots\;\;,
\end{equation}
We now come to the subleading terms. Let us introduce the notation
\begin{equation}
\bpic(20,0)(0,-2)\Cross{-2}{-2}{2}{2}\Line(0,0)(20,0)
\Cross{18}{-2}{22}{2}\Text(10,8)[c]{1}\epic\;\; =\;\; 
\bpic(20,0)(0,-2)\Cross{-2}{-2}{2}{2}\Line(0,0)(20,0)
\Cross{18}{-2}{22}{2}\epic
\;\;\;,\;\;\;
\bpic(20,0)(0,-2)\Cross{-2}{-2}{2}{2}\Line(0,0)(20,0)
\Cross{18}{-2}{22}{2}\Text(10,8)[c]{2}\epic\;\; = \;\;
\bpic(30,0)(0,-2)\Cross{-2}{-2}{2}{2}
\Line(0,0)(30,0)\Vix{15,0}\Cross{28}{-2}{32}{2}\epic
\;\;\;,\;\;\;
\bpic(20,0)(0,-2)\Cross{-2}{-2}{2}{2}\Line(0,0)(20,0)
\Cross{18}{-2}{22}{2}\Text(10,8)[c]{3}\epic\;\; =\;\; 
\bpic(40,0)(0,-2)\Cross{-2}{-2}{2}{2}\Line(0,0)(40,0)
\Vix{13.3,0}\Vix{26.6,0}\Cross{38}{-2}{42}{2}\epic
\;\;\;,
\end{equation}
so as to indicate the number of links between two vertices. The subleading
contributions are characterised by the additional presence of one
of the following diagrams:
$$
\bpic(20,4)(0,0)\Cross{0}{0}{4}{4}\Line(2,2)(18,2)
\Cross{16}{0}{20}{4}\Text(10,8)[c]{$q$}\epic
\;\;\;,\;\;\;
\bpic(40,20)(0,7)\BC(10,10){10}\BC(30,10){10}\Vix{20,10}
\Text(10,28)[c]{$q_1$}\Text(30,28)[c]{$q_2$}\epic
\;\;\;,\;\mbox{or}\;\;\;
\bpic(30,30)(0,10)\BC(15,15){15}\Line(0,15)(30,15)\Vix{0,15}\Vix{30,15}
\Text(15,38)[c]{$q_1$}\Text(15,23)[c]{$q_2$}\Text(15,8)[c]{$q_3$}\epic\;\;.$$
It is actually only the first of these graphs that is needed
for the purpose of computing $\rho_k$ to first order: the other
two will only contribute to subleading terms in $H_0(s)$ which
are not relevant to the order we are working in
here. The relevant contribution is, therefore
\begin{eqnarray}
\lefteqn{\rho_k(z;x_1,\ldots,x_k) = W_1(m) + \order{{1\over N}}\;\;,}\nl
W_1(m) & \equiv & 
\suml_{\stackrel{p_{1,2,\ldots}\ge0}{q>0}}
B(m;q,p_1,p_2,\ldots)\;
\bpic(25,4)(0,0)\Cross{0}{0}{4}{4}\Line(2,2)(23,2)
\Cross{21}{0}{25}{4}\Text(12.5,8)[c]{$q$}\epic\;
\left(\bpic(20,10)(0,7)\BC(10,10){10}
\Vix{10,20}
\epic\right)^{p_1}
\left(\bpic(20,10)(0,7)\BC(10,10){10}
\Vix{0,10}\Vix{20,10}
\epic\right)^{p_2}
\cdots\;\;,
\end{eqnarray}
with
\begin{equation}
m = q + p_1 + 2p_2 + 3p_3 + \cdots\;\;.
\end{equation}
The coefficient $B$ satisfies a recursion relation similar to that
of $A$:
\begin{eqnarray}
\lefteqn{B(m;q;p_1,p_2,\ldots) = 
2qB(m-1;q-1,p_1,p_2,\ldots)}\nl
& + & B(m-1;q,p_1-1,p_2,\ldots)\nl
& + & \suml_{k>0}2k(p_k+1)
B(m-1;q,p_1,p_2,\ldots,p_k+1,p_{k+1}-1,\ldots)\;\;.
\end{eqnarray}
By the same trick as above, we may solve this relation. The
symmetry factors for the loop diagrams are of course the same,
and the diagram 
$\bpic(25,4)(0,0)\Cross{0}{0}{4}{4}\Line(2,2)(23,2)
\Cross{21}{0}{25}{4}\Text(12.5,8)[c]{$q$}\epic$ has a symmetry
factor $2^{q-1}$. We therefore find
\begin{equation}
W_1(m) \equiv 
\suml_{\stackrel{p_{1,2,\ldots}\ge0}{q>0}}
{m!\over p_1!p_2!\cdots}\;\left({2^q\over2}\;
\bpic(25,4)(0,0)\Cross{0}{0}{4}{4}\Line(2,2)(23,2)
\Cross{21}{0}{25}{4}\Text(12.5,8)[c]{$q$}\epic\right)\;
\left({2^1\over2}\;\bpic(20,10)(0,7)\BC(10,10){10}
\Vix{10,20}
\epic\right)^{p_1}
\left({2^2\over4}\;\bpic(20,10)(0,7)\BC(10,10){10}
\Vix{0,10}\Vix{20,10}
\epic\right)^{p_2}
\cdots\;\;,
\end{equation}
In the above derivations, it may be noted that the various powers
of 2 arise from the fact that each link has two endpoints, while
the remaining symmetry factors are just those of diagrams with
topologically indistinguishable points ($1/2n$ for an
$n$-link loop, $1/2$ for the propagator diagram).

\subsection{The generating functions}
We can now write down immediately the leading form for
$G_0$ and $\rho_k$, by taking the
appropriate sums over $m$. To this leading order, we do not
have to worry about subleading terms in
$(N-k)\fal{m}$ and $(N-k)\fal{m-1}$. We have
\begin{eqnarray}
G_0(z) & \sim & \suml_{m\ge0}{z^m\over m!}
\langle D_N^m\rangle_0\;\; = \;\;
\suml_{m\ge0}{z^m\over m!}W_0(m)\nl
& = & \exp\left(
{2z\over2}\;\bpic(20,20)(0,7)\BC(10,10){10}
\Vix{10,20}
\epic
\;\;+\;\; 
{(2z)^2\over4}\;\bpic(20,20)(0,7)\BC(10,10){10}
\Vix{0,10}\Vix{20,10}
\epic
\;\;+\;\; 
{(2z)^3\over6}\;\bpic(20,20)(0,7)\BC(10,10){10}
\Vix{10,20}\Vix{3.5,3.5}\Vix{16.5,3.5}
\epic
\;\;+\cdots \right)\;\;,
\end{eqnarray} 
and
\begin{equation}
\rho_k(z;x_1,\ldots,x_k)\;\;\sim\;\;\suml_{m\ge1}{z^m\over m!}W_1(m)
\;\;=\;\;{1\over2}\suml_{q>0}
\left((2z)^q\vphantom{{A^A\over A^A}}
\bpic(25,4)(0,0)\Cross{0}{0}{4}{4}\Line(2,2)(23,2)
\Cross{21}{0}{25}{4}\Text(12.5,8)[c]{$q$}\epic\right)\;G_0(z)\;\;.
\end{equation}
To evaluate the various diagrams we employ the functions $\beta$.
From the following propagator diagrams, with endpoints $x$ and $y$:
\begin{eqnarray}
\bpic(25,0)(0,0)
\Cross{-2}{-2}{2}{2}\Line(0,0)(25,0)\Cross{23}{-2}{27}{2}
\Text(12.5,8)[c]{$1$}\epic & = & 
\beta_1(x,y)\;\;,\nl
\bpic(25,0)(0,0)
\Cross{-2}{-2}{2}{2}\Line(0,0)(25,0)\Cross{23}{-2}{27}{2}
\Text(12.5,8)[c]{$2$}\epic & = & 
\intk dz\;\beta_1(x,z)\beta_1(z,y) = \beta_2(x,y)\;\;,\nl
\bpic(25,0)(0,0)
\Cross{-2}{-2}{2}{2}\Line(0,0)(25,0)\Cross{23}{-2}{27}{2}
\Text(12.5,8)[c]{$3$}\epic & = & 
\intk dz\;\beta_2(x,z)\beta_1(z,y) = \beta_3(x,y)\;\;,\nl
& \vdots & \nl
\bpic(25,0)(0,0)
\Cross{-2}{-2}{2}{2}\Line(0,0)(25,0)\Cross{23}{-2}{27}{2}
\Text(12.5,8)[c]{$q$}\epic & = & \beta_q(x,y)\;\;=\;\;
\suml_{\vec{n}>0}(\si_{\vec{n}})^{2q}\;u_{\vec{n}}(x)u_{\vec{n}}(y)\;\;,
\end{eqnarray} 
we find the following representations:
\begin{eqnarray}
\phi(z;x,y) & \equiv &
\suml_{q\ge1}(2z)^q\;\;
\bpic(25,0)(0,0)
\Cross{-2}{-2}{2}{2}\Line(0,0)(25,0)\Cross{23}{-2}{27}{2}
\Text(12.5,8)[c]{$q$}\epic \;\;= \;\;
\suml_{\vec{n}>0}{2z\si_{\vec{n}}^2\over1-2z\si_{\vec{n}}^2}\;u_{\vec{n}}(x)u_{\vec{n}}(y)\;\;,
\\
\log G_0(z) & = & -{1\over2}\suml_{\vec{n}>0}
\log\left(1-2z\si_{\vec{n}}^2\right)\;\;,
\label{logg0definition}
\end{eqnarray}
and
\begin{equation}
\rho_k(z;x_1,\ldots,x_k) = {1\over2}\suml_{i\ne j}
\phi(z;x_i,x_j) G_0(z)\;\;,
\label{rhodefinition}
\end{equation}
with the indices $i$ and $j$ running from 1 to $k$.
For the case $k=2$ in which we are interested, this
specializes to
\begin{equation}
\rho_2(z;x_1,x_2) = \phi(z;x_1-x_2)G_0(z)\;\;.
\end{equation}
In the language of field theory, 
the function $\rho_2(z;,x_1,x_2)/N$ may be recognized as 
the Dyson-summed, or {\em dressed}, two-point Green's function,
including vacuum diagrams $G_0(z)$.
Here $G_0(z)$ plays the r\^{o}le of the path integral in the free-field
approximation.

\section{Conclusions}
In this first paper we have addressed the question of how, given
information on the uniformity property of a particular point set
$X_N$ in terms of a discrepancy (defined in 
\eqn{discrepancydefinition}), we may hope to improve on the
error estimate when we apply the point set $X_N$ in numerical
integration. To this end, we need the function $F_2(s;x_1,x_2)$
which is defined as the ratio of two functions, 
$R_2(s;x_1,x_2)$ and $H_0(s)$, which are themselves again given
as the Fourier transforms of $\rho_k(z;x_1,x_2)$ and $G_0(z)$,
respectively. We have developed a diagrammatic approach that
allows us to systematically compute a series expansion for 
these two objects in powers of $1/N$. We have derived explicit
expressions for the leading term in these expansions: they are
given by \eqn{rhodefinition} and \eqn{logg0definition}, respectively.

Of course, the above discussion has been purely formal.
We have been able to point out interesting parallels between
our results and properties of a classical field theory; but
all this will be moot unless we can, in fact, turn what we
have learned into an expression for $F_2(s;x_1,x_2)$ that
is explicit and simple to evaluate. This will be
the subject of further publications in this series.

Apart from the explicit expressions for $\langle D_N\rangle_k$
and $\langle D_N^2\rangle_k$, our results have been strictly
limited to the leading terms, that are independent of $N$. 
Since any numerical integration usually employs a large value
for $N$ anyway, we feel that we are justified in this.
However, there are issues for which a further expansion ought to
be useful. For instance, it would be interesting to see how
far we could approximate a lower bound on $D_N$ \`{a} la Roth,
by either computing further terms in $F_2$ and putting its value
to $N$, or by trying to establish bounds on that value for $s$
where $H_0(s)$ vanishes, for finite $N$. These issues are
beyond our scope at this moment, but we feel that we have at
least laid some of the groundwork in this paper.

\section*{Appendix A: Expansion of falling powers}
In this appendix we give a simple derivation of the asymptotic
expansion of the quantity $(N-k)\fal{m}$. It is based upon
the binomial-theorem fact
\begin{equation}
\suml_{m\ge0}{1\over m!}a\fal{m}x^m = (1+x)^a\;\;.
\end{equation}
We therefore have
\begin{equation}
\suml_{m\ge0}{1\over m!}(N-k)\fal{m}\left({x\over N}\right)^m =
\left(1+{x\over N}\right)^{(N-k)}\;\;.
\end{equation}
Now, one the one hand, we have
\begin{equation}
\left(1+{x\over N}\right)^N = 
e^x\left(1 - {x^2\over2N} + {x^3\over3N^2} +
{x^4\over8N^2} + \order{N^{-3}}\right)\;\;,
\label{expnsr}
\end{equation}
and, on the other hand,
\begin{equation}
\left(1+{x\over N}\right)^{-k} = 
1 - {kx\over N} + {k(k+1)x^2\over 2N^2} + \order{N^{-3}}\;\;.
\label{inversandum}
\end{equation}
Multiplying the expansions (\ref{inversandum}) and (\ref{expnsr}), and carefully keeping track of the coefficient of $x^m$, we then find
\begin{eqnarray}
{(N-k)\fal{m}\over N^m} & =&  1 
- {1\over N}\left[{1\over2}m\fal{2} + km\right]\nl
& & + {1\over N^2}\left[{1\over8}m\fal{4} + {1\over3}m\fal{3} 
+ {1\over2}km\fal{3} + {1\over2}k(k+1)m\fal{2}\right]\nl  
& & + \order{N^{-3}}\;\;,
\end{eqnarray}
and, of course, also the subleading expressions
\begin{eqnarray}
(N-k)\fal{m-1}/N^m & = & {1\over N} 
- {1\over N^2}\left[{1\over2}m\fal{2} + (k-1)(m-1)\right]
+ \order{N^{-3}}\;\;,\nl
(N-k)\fal{m-2}/N^m & = & {1\over N^2} + \order{N^{-3}}\;\;.
\end{eqnarray}
Obviously, this expansion can be continued {\it ad nauseam}; moreover,
it also provides a way of computing complicated 
constraint sums of products like
$$
\suml_{k_1,k_2,\ldots,k_p=1}^{N-k}k_1k_2\cdots k_p\;\;,
$$
with the constraints that all indices must be different.

\section*{Appendix B: an illustrative model}
In this appendix we show how error improvement arises in a very
simple one-dimensional model. 
We assume that the functions to be integrated
have only modes up to $n$, and that those modes have equal strength. 
That is, we take
\begin{eqnarray}
\si_k^2 = 1/2 & , & k=1,2,\ldots,2n\;\;,\nl
\si_k^2 = 0   & , & k>2n\;\;.
\end{eqnarray}
For this extremely simple case, we have
\begin{equation}
G_0(z) = {1\over(1-z)^n}\;\;\;,\;\;\;
\phi(z;x) = {2z\over1-z}\suml_{k=1}^n\cos(2\pi kx)\;\;.
\end{equation}
We may easily compute the Laplace/Fourier transforms, and find
\begin{equation}
H_0(s) = {1\over(n-1)!}s^{n-1}e^{-s}\;\;\;,\;\;\;
F_2(s;x) = 2\left(1-{s\over n}\right)\suml_{k=1}^n\cos(2\pi kx)\;\;.
\end{equation}
The average discrepancy is $\langle s \rangle=n$ for truly random points.
A typical integrand is, in this model, of the form
\begin{equation}
f(x) = \suml_{k=1}^n v_k\,\cos(2\pi k(x+\alpha_k))\;\;,
\end{equation}
with arbitrary $v_k$ and $\alpha_k$. We immediately find
\begin{equation}
J_1 = 0\;\;\;,\;\;\;J_2 = {1\over2}\suml_{k=1}^nv_k^2\;\;,
\end{equation}
and
\begin{equation}
\intk dx_1dx_2\;f(x_1)f(x_2)F_2(s;x_1-x_2) =
\left(1 - {s\over n}\right){1\over2}\suml_{k=1}^n v_k^2\;\;.
\end{equation}
We conclude that, with the inclusion of the two-point correlation 
function $F_2$, the straightforward \mc\ error estimate is changed
into the \qmc\ one, as follows:
\begin{eqnarray}
\lefteqn{\langle\eta^2\rangle_{\mbox{{\scriptsize\mc}}} 
\;\;\rightarrow}\nl
& \rightarrow &
\langle\eta^2\rangle_{\mbox{{\scriptsize\qmc}}}\;=\; 
{s\over \langle s \rangle}\langle\eta^2\rangle_{\mbox{{\scriptsize\mc}}}
\;+\;
\order{{1\over N^2}}
\;.
\end{eqnarray}
We see that the error is improved
if the actual discrepancy is small compared to its expected value.


\end{document}